\documentclass[12pt,preprint,apj]{emulateapj}

\newcommand{\kms}{{km s$^{-1}$}~}

\newcommand{\kmsMpc}{{km s$^{-1}$ Mpc$^{-1}$}~}
\usepackage{color}

\shorttitle{A Distance to NGC 4993}
\shortauthors{Lee et al.}

\usepackage{multirow}

\begin{document}


\title{
A Globular Cluster Luminosity Function Distance to NGC 4993 Hosting a Binary Neutron Star Merger GW170817/GRB 170817A
}

\email{mglee@astro.snu.ac.kr}

\author{Myung Gyoon Lee, Jisu Kang, and Myungshin Im}

\affiliation{Astronomy Program, Department of Physics and Astronomy, Seoul National University, Gwanak-gu, Seoul 08826, Korea}
\begin{abstract}
{NGC 4993 hosts a binary neutron star merger emitting gravitational waves and electromagnetic waves, GW170817/GRB 170817A.
The distance to this galaxy is not well established. 
We select the globular cluster candidates from the Hubble Space Telescope/ACS F606W images of NGC 4993 in the archive, using the structural parameters of the detected sources.
The radial number density distribution of these candidates shows a significant central concentration around the galaxy center at the galactocentric distance $r<50''$, showing that they are mostly the members of NGC 4993.
Also the luminosity function of these candidates is fit well by a Gaussian function.
Therefore the selected candidates at $r<50''$ are mostly considered to be globular clusters in NGC 4993.
We derive an extinction-corrected turnover Vega magnitude in the luminosity function of the globular clusters at $20''<r<50''$, F606W (max)$_0= 25.36\pm0.08$ ($V_0 =25.52\pm0.11$)} mag.
Adopting the calibration of the turnover magnitudes of the globular clusters, $M_V({\rm max})=-7.58\pm0.11$, 
we derive a distance to NGC 4993, $d=41.65\pm3.00$ Mpc ($(m-M)_0=33.10\pm0.16$).
The systematic error of this method can be as large as $\pm0.3$ mag.
This value is consistent with the previous distance estimates based on the fundamental plane relation and the gravitational wave method in the literature.
The distance in this study can be used to constrain the values of the parameters including the inclination angle of the binary system in the models of gravitational wave analysis. 
\end{abstract}

\section{Introduction}

Both gravitational waves and electromagnetic waves  were detected  in NGC 4993 with a time delay of 2s (GW170817/GRB 170817A) on 2017 August 17, and their origin is believed to be the merger of a binary neutron star system located at $10''$ from the galaxy center \citep{abb17,gol17,sav17,cou17}. This is the first system that was detected directly in both gravitational waves and electromagnetic waves.  
Since then NGC 4993 hosting 
GW170817/GRB 170817A 
has become a target of numerous studies.
The distance to NGC 4993 is a critical information to investigate the properties of the progenitor of GW170817/GRB 170817A \citep{sma17,tan17,tanv17,abb17b}.

However, the distance to NGC 4993 is not well established.
Previous estimates of the distance to NGC 4993 are based on the fundamental plane relation or the gravitational wave method, 
ranging from 38 Mpc to 44 Mpc \citep{im17,hjo17,abb17,abb17b}. 
Indirect distance indicators such as redshifts or the group membership were also used to provide a value of around 40 Mpc for the distance to NGC 4993 \citep{hjo17}. 

The fundamental plane relation provides angular diameter distances for elliptical galaxies or bulges of spiral galaxies \citep{dre87,djo87}.
However, it is known to suffer from metallicity and environment effects,  and  its zeropoint is not well established \citep{jor96,hjo97,ber03,lab10}.

On the other hand, the gravitational wave method is based on the models of binary systems with gravitational wave emission, being independent of the cosmic distance ladder \citep{sch86,abb17,abb17b}. But, the distance determination in this method depends on model parameters such as the inclination angle of the binary system.

NGC 4993 provides a unique chance to compare the gravitational wave distances and the cosmic ladder distances, and to constrain the models of binary systems with gravitational wave emission. 
However, there are, to date, no published luminosity distances to NGC 4993 based on the 
standard candles.

NGC 4993 is a member of the NGC 4993 group,
which is a subgroup of ESO 508 
at the distance of about 40 Mpc \citep{sak00,kou17,hjo17}. 
It is a lenticular galaxy with a morphological type of (R')SAB(rs)0-\citep{dev91}. It is also listed as an elliptical galaxy with two shells  
\citep{mal83}.
NGC 4993 shows some features of recent merger including dust lanes in the central region 
\citep{im17,lev17}.
The central velocity dispersion of NGC 4993 is as large as $\sigma_v \approx 170$ \kms\citep{im17,hjo17}, indicating that NGC 4993 is a massive galaxy.
From its early-type morphology and large central velocity dispersion, it is expected that NGC 4993 may host a significant population of globular clusters \citep{hud14}. 

In this Letter, we 
determine a luminosity distance to NGC 4993, using the globular cluster luminosity function (GCLF). The turnover magnitude of the GCLF 
is one of the well-known standard candles, and has been applied out to Coma galaxies \citep{har01,ric03,dic06,rej12,lee16}.
%

\begin{figure*} 
	\centering
	\includegraphics[scale=0.7]{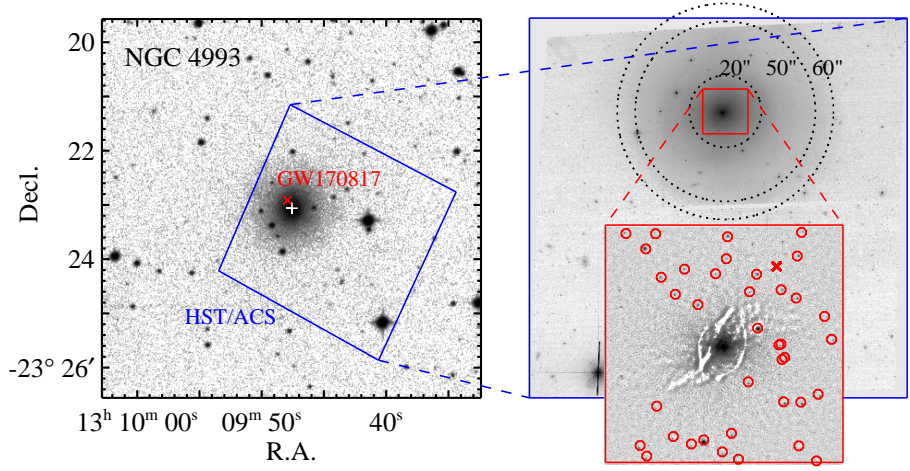} 
	\caption{(Left) The footprint of the HST/ACS field of NGC 4993 on the digitized sky survey map. The position of GW170817 is marked by a red cross close to the galaxy center. 
		(Right) A gray-scale map of the F606W image of NGC 4993. The large circles represent the boundary for the galaxy region 
		($r<50''$) and the background region ($r>60''$) defined for our analysis, respectively.
		The lower box is a zoomed-in image of the central $25'' \times 25''$ region, from which galaxy light is subtracted. Globular cluster candidates are marked by red circles.
	}
	\label{fig_finder}
\end{figure*}

\section{Data and Data Reduction}


We used ACS F606W images of NGC 4993 in the Hubble Space Telescope (HST) archive
(PI: Andrea Bellini, ID: 14840). These images were obtained on April 28, 2017.
There are only one band images of NGC 4993 available in the archive at the time of this study. 
 There are two images each of which has an exposure time of 348 s, and these were combined using AstroDrizzle by the STScI team. 
The image scale of the combined image is  0\farcs05 per pixel, and the FWHM value of the point sources 
is 2.0 pixels.

For better source detection, we subtracted the contribution of galaxy light of NGC 4993 
from the original image. 
We smoothed the original image using IRAF/rmedian (with ring median filter radius = 10 pixels), and subtracted the smoothed image from the original image. 

{\color{blue}\bf Figure \ref{fig_finder}} shows 
the location of the ACS field on the digitized survey map, and a gray-scale map of the F606W image of NGC 4993.
We also display a zoomed-in  image of the central $25''\times 25''$ region of NGC 4993, from which the galaxy light was subtracted. The zoomed-in image shows clearly  interesting ring-like features of dust lanes around the galaxy center,   
indicating that NGC 4993 might have had a merger recently. 
The counterpart of GW170817/GRB 170817A is located at $\sim10''$ (2 kpc) 
in the north-east from the galaxy center.
The point-like sources in the central region marked by red circles are globular cluster candidates.

We detected sources with detection threshold of $2\sigma$ 
that is estimated from the local background on the image where galaxy light was subtracted, using Source Extractor \citep{ber96}.
%
We obtained aperture photometry of the detected sources with a small aperture for radius of 2.5 pixels = $0\farcs125$, F606W($R<2.5$ pixel), as a first step to select the globular cluster candidates among the detected sources. 
We use the Vega system for the magnitudes of the sources.


\subsection{Globular Cluster Candidate Selection}

\begin{figure*} 
	\centering
	\includegraphics[scale=0.7]{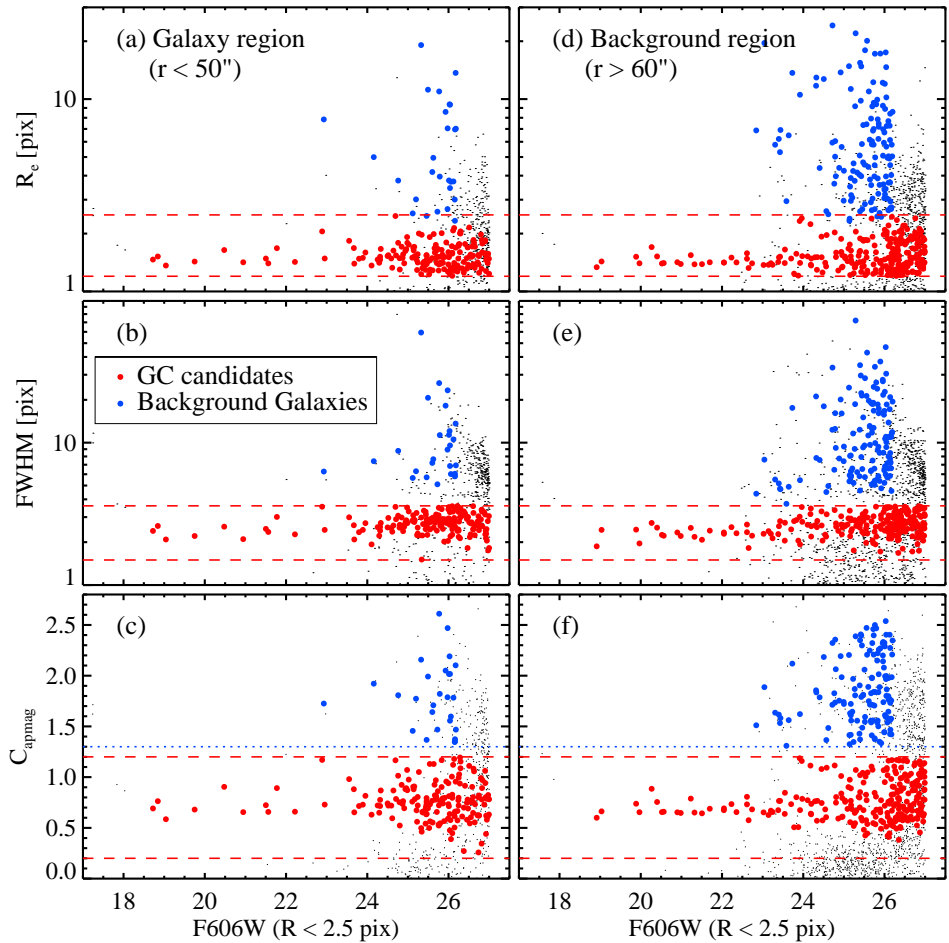} 
	\caption{Structural parameters vs. F606W magnitudes ($R<$2.5 pixel) for the detected sources in the galaxy region 
		(Left) and the background region 
		(Right).
		Red and blue circles represent the globular cluster candidates in NGC 4993 and the background galaxy candidates, respectively.
		Black dots are the detected sources that are not selected either as globular cluster candidates nor background galaxies. 
		Red dashed lines mark the boundary for globular cluster candidate selection,
		and blue dotted lines mark the boundary for background galaxy candidate selection.}
	\label{fig_param}
\end{figure*}

\begin{figure} 
	\centering
	\includegraphics[scale=0.48]{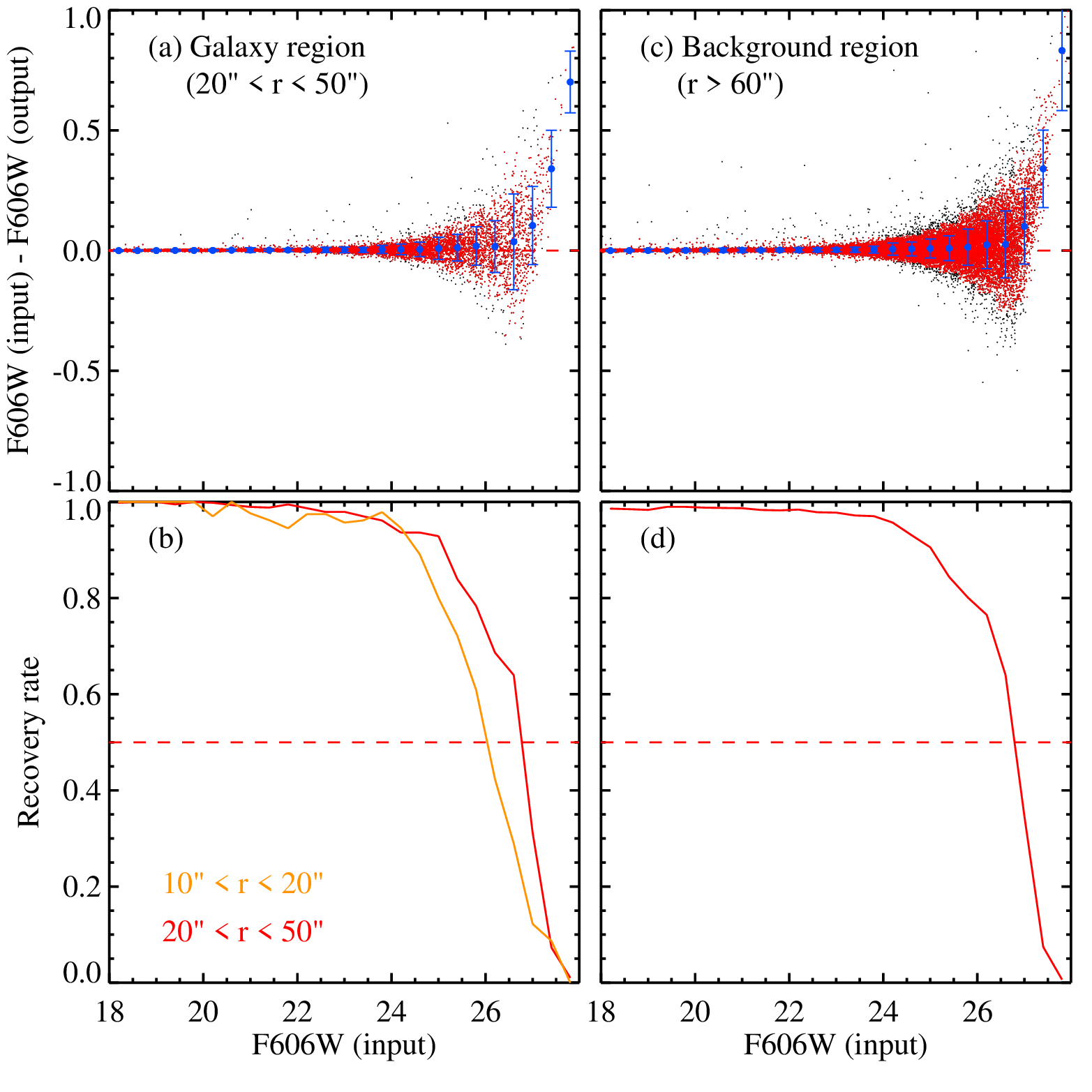}
	\caption{Completeness test results for globular cluster-like sources (with FWHM=2.2 pixels)  in the galaxy region (Left) and the background region ($r>60''$) (Right). (Top) Magnitude difference (input magnitudes minus output magnitudes) vs. input F606W magnitudes. Blue circles with errorbar denote the mean values of the data with 
		$\pm1 \sigma$. (Bottom) Completeness vs. input F606W magnitudes.  In (b), the 
		orange line is for $10''<r<20''$ and the red one for $20''<r<50''$.
	}
	\label{fig_comptest}
\end{figure}

The image scale at the distance of 40 Mpc is 193 pc arcsec$^{-1}$ so one pixel ($0\farcs05$) in the ACS image corresponds about 10 pc. Therefore it is expected that typical globular clusters (with effective radii of 2--3 pc) at the distance of NGC 4993 would appear as point sources or slightly extended sources in the HST/ACS images.
We selected the globular cluster candidates among the detected sources with Source Extractor parameter 
FLAGS $<4$, using three criteria based on structural parameters: two Source Extractor parameters (FLUX\_RADIUS (effective radius $R_{\rm eff}$) and FWHM), 
and one concentration parameter we derived. 
We derived the concentration parameter of the sources using the difference in the aperture magnitudes with small and large aperture (with radii of 1.5 pixels and 5 pixels, respectively):
$C_{\rm apmag} = {\rm mag(1.5~ pixel) - mag (5~ pixel)}$.
This parameter is known to be very effective and reliable for distinguishing point-like sources from extended sources like background galaxies and has been used for selecting globular cluster candidates in other galaxies \citep{lee16}. 

For the following analysis, we divided the HST field into two regions using the boundary as described in Section 3.1: 
the galaxy region at $r<50''$ and the background region at $r>60''$, as marked
in {\color{blue}\bf Figure \ref{fig_finder}}. 

{\color{blue}\bf Figure \ref{fig_param}} displays the values of these parameters versus F606W($R<2.5$ pixel) magnitudes of the detected sources 
in the galaxy region and the background region. 
Note that there is a narrow horizontal sequence of bright sources in each diagram at
$R_{\rm eff} \approx 1.5$ pixels, 
FWHM $\approx 2$ pixels,
and
$C_{\rm apmag}\approx 0.6$.
The detected sources with much larger values of these parameters are 
significantly extended sources like  background galaxies.
It is expected that the globular clusters in NGC 4993 are located around the narrow horizontal sequence in each diagram.
Therefore we used the following criteria for selecting the globular cluster candidates:
$18.5<$F606W(2.5 pixel)$<27.0$ mag,
$1.2<R_{\rm eff}<2.5$ pixels, 
$1.5<$FHWM$<3.6$ pixels,
and
$0.2<C_{\rm apmag}<1.2$.
We set the lower limit of F606W magnitude to select sources with photometric errors smaller than 0.1 mag.
This value is about one magnitude brighter than the 2$\sigma$ detection limit. 
For comparison we also selected extended sources with F606W(2.5 pixel)$<26.2$ mag and
$C_{\rm apmag}>1.3$, which are mostly background galaxies.
We inspected visually the images of these globular cluster candidates, removing artefacts, galaxy-like sources, sources close to the image edge, and faint sources close to bright stars. 

We estimated the aperture correction for the difference between $R=0\farcs125$ magnitudes and $R=0\farcs5$ magnitudes, using the bright point-like sources (globular cluster-like sources). We derived a relation between aperture correction (APC) and $C_{\rm apmag}$: APC$(0\farcs125 - 0\farcs5 )= 0.31 C_{\rm apmag} + 0.04$ with rms=0.005, 
from the photometry of bright sources in the image.
We applied this correction to the magnitudes with $R=0\farcs125$ of the sources. 
Finally we derived the total magnitudes of the sources, using aperture correction for $0\farcs5$ to infinity radius for F606W given in \citet{sir05}, 0.088 mag. 

\subsection{Completeness Test}

We estimated the completeness of our selection of the globular cluster candidates, using an artificial source experiment. We generated the images of artificial sources
with structural parameters  
similar to the slightly extended sources 
(FWHM = 2.2 pixel), 
using IRAF/ARTDATA. 
We assumed that the magnitude distribution of the artificial sources is uniform from 18 to 28 mag. We added 1000 artificial sources onto the original image to generate one test image. We repeated this procedure 100 times, generating 100 test images. 
 The total number of added artificial sources in 100 test images is 100,000. 
We analysed these images using the same procedure as done for the original image. 
We derived the completeness from the number ratio of the recovered sources and the added sources.

{\color{blue}\bf Figure \ref{fig_comptest}} displays the results of the completeness test for the galaxy region (for $10''<r<20''$ and $20''<r<50''$) and background region.
%
The completeness is lower in the central region at $r<20''$, while it varies only slightly depending on the galactocentric distance in the outer region at  $r>20''$.
Thus we use only the outer region at $r=20''-50''$ 
to derive the luminosity function of the globular clusters.
The average completeness for $r=20''-50''$ 
is almost 
100\% for the bright magnitudes, and decreases in the fainter magnitudes, reaching 50\% at
F606W$\approx 26.7$ mag. 
Using this information we corrected the incompleteness in the luminosity function of the globular cluster candidates.
The mean difference between the input magnitudes and the output magnitudes is $\sim$0.01 mag for F606W$<27.0$ mag. 
We correct it for 
the GCLF turnover magnitude.

\begin{figure*} 
	\centering
	\includegraphics[scale=0.7]{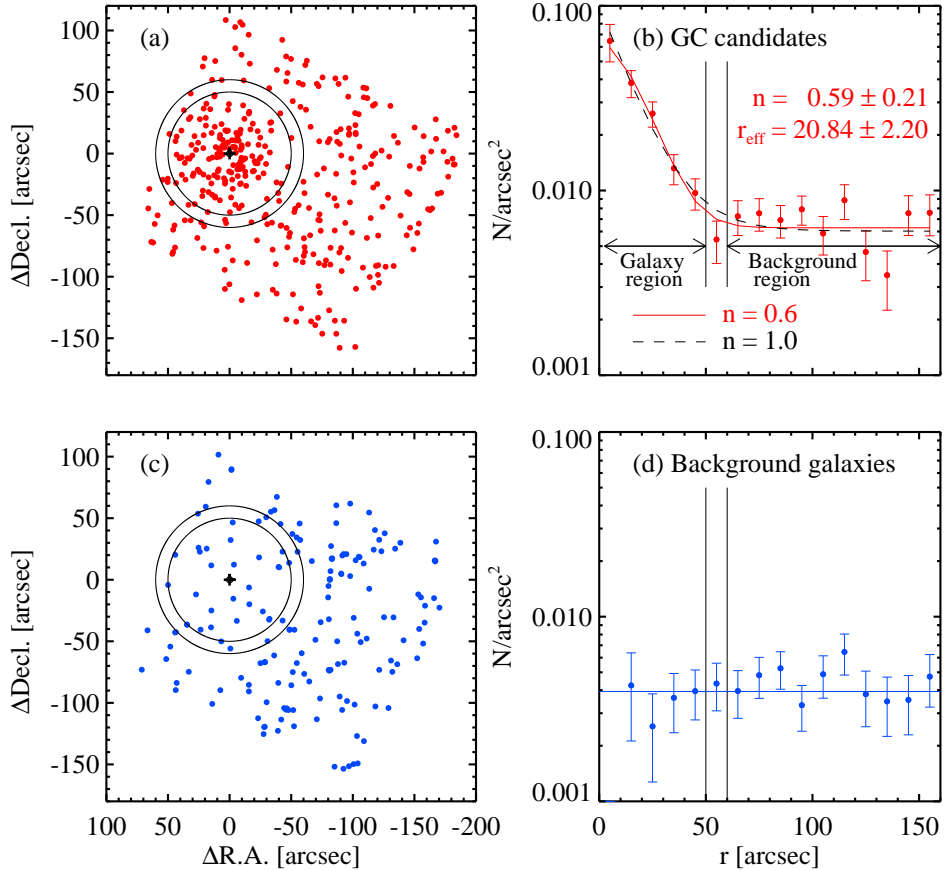} 
	\caption{Spatial distributions (Left)  and radial number density profiles  (Right)  of the globular cluster candidates with F606W $<26.4$ mag (red circles) and the background galaxy candidates with F606W($R<2.5$ pix) $<26.2$ mag
		(blue circles). Large circles and vertical lines mark the boundary for the galaxy 
		and background regions. 
		The solid line in (b) represents a fitting with a S{\'e}rsic law, in comparison with the case for $n=1$ (dashed line).
	}
	\label{fig_spat}
\end{figure*}

\section{Results}

\subsection{Spatial and Radial Distributions of the Globular Clusters}

{\color{blue}\bf Figure \ref{fig_spat}} (Top) displays the spatial distribution and radial number density profile of the globular cluster candidates 
with F606W$<26.4$ mag
in the ACS field of NGC 4993.
For comparison, {\color{blue}\bf Figure \ref{fig_spat}} (Bottom) displays 
 same figures but for the extended sources 
with F606W(2.5 pixel)$<26.2$ mag. 
We used annular apertures with a bin size of $10''$ to derive the radial number density profile for $r<160''$.
The spatial and radial number densities of the globular cluster candidates show a significant central concentration at $r<50''$. Their radial number density decreases as the galactocentric distance increases, and becomes almost flat at $r>60''$. This shows that the selected globular cluster candidates at $r<50''$ are mostly the members of NGC 4993.
NGC 4993, with a morphological type of lenticular/elliptical galaxies, do not show any 
star-forming regions 
so that the selected candidates are dominated by genuine globular clusters rather than by young supergiant stars in NGC 4993.   
Most of the sources in the outer region at $r>60''$ are probably foreground stars or unresolved background galaxies. 
On the other hand, the spatial and radial number densities of the extended sources show almost uniform distributions,
indicating that the extended sources are mainly background galaxies in the direction of NGC 4993.


We fit the radial number density profile of the globular cluster candidates for $r<160''$ with a S{\'e}rsic law including the background level \citep{ser63,gra05}. The data for $r>60''$ represent mainly the background level.
The parameter values we obtained are $n=0.59\pm0.21$, ${r_{\rm eff}=20\farcs84\pm2.20}$, and the background level of $\Sigma = 0.006$ sources arcsec$^{-1}$. 
To test the binning effect, we performed fitting for different bin sizes of $5''$, $8''$, and $15''$, obtaining $n=0.60\pm0.21$, $n=0.61\pm0.21$, and $n=0.57\pm0.19$, respectively. These values all agree within the errors, showing the binning effect is negligible.
We also plotted an exponential profile with $n=1$ as a reference (the dashed line). 
The value of the S{\'e}rsic' index we observe is not relevant to the main results for distance estimation.
The effective radius of the globular cluster system is larger than that of the galaxy light, $r_{\rm eff,star}=15''-16''$ \citep{im17,hjo17}. 

\subsection{Luminosity Function of the Globular Clusters}

\begin{figure} 
	\centering
	\includegraphics[scale=0.95]{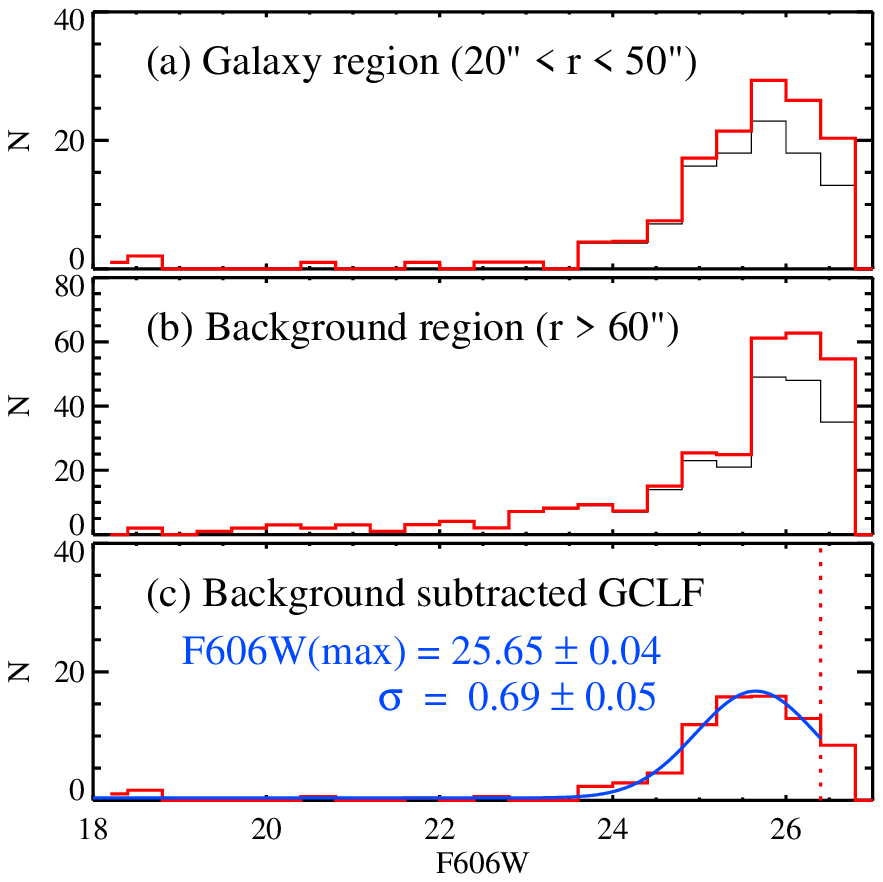}
	\caption{(a, b) The luminosity functions of the globular cluster candidates in the galaxy 
		and background regions.
		Black and red histograms represent the luminosity functions before and after completeness correction, respectively.
		(c) The GCLF for NGC 4993,
		corrected for background contamination.
		The dashed line  shows a Gaussian fitting for the range of F606W$<26.4$ mag.
	}
	\label{fig_gclf}
\end{figure}

We  derived the luminosity function of the globular cluster candidates in NGC 4993, counting the sources in the galaxy region at 
$20''<r<50''$. We excluded the central region at $r<20''$ to avoid any extinction effect due to dust lanes. 
We estimated the contribution of the foreground or background sources, counting the sources in the background region at $r>60''$. 

In {\color{blue}\bf Figure \ref{fig_gclf}}(a) and (b) we plot the luminosity functions of the globular cluster candidates in the galaxy region and background region, respectively, before and after completeness correction.
We display the background-subtracted  luminosity function for the galaxy region in {\color{blue}\bf Figure \ref{fig_gclf}}(c). 
The luminosity function of the selected globular cluster candidates in the galaxy region in {\color{blue}\bf Figure \ref{fig_gclf}}(c) appears to be approximately Gaussian, showing a turnover (peak) at F606W$\approx 25.6$ mag.
Both of the spatial distribution with central concentration and the luminosity function  with a Gaussian form show that the selected globular cluster candidates at $r<50''$ are, indeed, mostly genuine globular clusters in NGC 4993.
  
We fit the background-subtracted luminosity function for the galaxy region,
using a Gaussian function. We use the magnitude range for fitting, F606W$<26.4$ mag where completeness is higher than 60\%.
The derived values of the turnover magnitude and width are
F606W${\rm (max)} = 25.65\pm0.04$ mag and $\sigma = 0.69\pm0.05$, respectively. 

We tested the effect of bin sizes, cutoff magnitudes, and galactocentric radii using the Jackknife resampling method.
The parameter values we tested are bin sizes of 0.2 and 0.4 mag, cutoff magnitudes of F606W = 26.4 and 26.8 mag, and galactocentric radii of $10''<r<50''$ and $20''<r<50''$.
%
In summary the values of the turnover magnitude are changed by $\pm0.04$ mag, if we use different bin sizes, cutoff magnitudes, and galactocentric radii. We included it in the error calculation.

Then, we derive 
F606W${\rm (max)}_0 = 25.36\pm0.08$ 
 mag after correction for foreground extinction 
($A_{\rm F606W}=0.305\pm0.049$)\citep{sch11}  
and 
the magnitude difference (F606W(input) - F606W(output) = $0.01\pm0.00$). 
We transform F606W turnover magnitudes of NGC 4993 globular clusters to the $V$ magnitudes in the Johnson-Cousins system using the information in \citet{sir05} (in their Table 22).
Adopting $(V-I)=1.0\pm0.05$ as a mean color of the globular clusters \citep{lee16},
we obtain 
$V{\rm (max)}_0 = 25.52\pm0.11$ mag.

 \subsection{GCLF Distance Estimation for NGC 4993}

The turnover magnitude of the GCLFs 
has been used as a distance indicator for nearby galaxies with various types
(see \citet{har01,ric03,dic06,rej12,lee16} and references therein). 
We adopt the calibration of the turnover $V$-band magnitudes based on the sample of the selected 100 globular clusters in the Milky Way given by \citet{dic06}: 
$M_V {\rm (max)} = -7.58\pm0.11$ mag. 
This sample consists of the globular clusters with 
relatively 
low reddening (with $E(B-V)<1.0$) 
at the outer region of the Milky Way Galaxy ($2<r_{GC}<35$ kpc). It includes both metal-poor and metal-rich globular clusters, but it is dominated by metal-poor globular clusters. 
Since we do not have any information on the colors of the globular cluster candidates in NGC 4993, we apply the calibration based on the combined sample of metal-poor and metal-rich globular clusters.

Applying this calibration to the measured turnover magnitude of the NGC 4993 globular clusters, we derive a distance to NGC 4993: 
$(m-M)_0 =33.10\pm0.11{\rm (ran)}\pm0.11{\rm (sys)} = 33.10\pm0.16$ ($d=41.65\pm3.00$ Mpc). 

\subsection{Uncertainties of the GCLF Method}

As a systematic error for our distance modulus estimate, we quote the calibration error of the turnover magnitude, $\pm0.11$ mag, given by \citet{dic06}. However,
the real systematic error must be larger than this value, considering the intrinsic uncertainties of the GCLF method. 

\citet{rej12} presented an extensive review on the uncertainties of the GCLF method,
summarizing that the total error of the estimated distance modulus for a galaxy can amount to $\sim$0.3 mag. The sources of errors included in \citet{rej12} are
1) uncertainty in the primary calibrators,
2) intrinsic dispersion of the turnover magnitudes and its dependence on the globular cluster sample,
3) dependence of the turnover magnitudes on environment and dynamical evolution,
4) dependence of the turnover magnitudes on the Hubble type or galaxy luminosity,
5) corrections due to a different metallicity of the target galaxies and the calibrator globular cluster sample,
and 6) the uncertainty in the age of the globular cluster sample.
Precise quantitative estimation of the errors for each of these sources is not easy so \citet{rej12} presents only an approximate value for the total error.

We discuss the sources relevant to NGC 4993 among the error sources in \citet{rej12}.   NGC 4993 is located in a loose group, and the Milky Way is in the Local Group. This implies that the effects of environment and dynamical evolution on the turnover magnitudes for NGC 4993 and the Milky Way are similar, leading to a small error.  NGC 4993 is as luminous as the Milky Way. Therefore the effects of galaxy luminosity on the turnover magnitudes 
  and corrections due to a different metallicity of the target galaxies and the calibrator globular cluster sample
are similar for NGC 4993 and the Milky Way , indicating the error due to these factors must be small.
Considering these, the total error for the case of NGC 4993 must be smaller than the value, $\sim$0.3 mag, given in \citet{rej12}. 

In addition we discuss two more uncertainties of the GCLF method applied to NGC 4993 in this study.
First, we checked the effect of reddening on the calibration of the GCLF for the Milky Way sample.
The Milky Way globular cluster sample we adopted is one with less extinction of the two sample cases that \citet{dic06} prepared using the Harris 2003 version catalog \citep{har96}. The globular clusters with higher extinction are mainly located in the direction of the Milky Way. Thus the selected sample with less extinction (also located in the outer region) in the Milky Way would be more similar to the case for the globular clusters in other galaxies we observe, than the entire sample. 
We checked the effect of reddening on the turnover magnitudes using the updated catalog of the Milky Way Globular Cluster (Harris 2010 version). We derived the turnover magnitudes for the samples at $2<R<35$ kpc with varying $E(B-V)$ limits (there are no GCs with $0.8<E(B-V)<1.0$, so the real range is $E(B-V)<0.8$): 
$M_V=-7.59\pm0.08$ ($\sigma=0.95\pm0.10$) for $E(B-V)<0.8$ (N=103), which is very similar to the value given by \citet{dic06},  
$M_V=-7.62\pm0.08$ ($\sigma=0.96\pm0.10$) for $E(B-V)<0.5$ (N=85), and
$M_V=-7.68\pm0.12$ ($\sigma=1.01\pm0.14$) for $E(B-V)<0.3$ (N=66).
These values agree within the errors, implying that the errors due to the reddening selection are not significant. 

Second, we checked the effect of subpopulations of globular clusters.
In general, massive galaxies host two subpopulations of globular clusters, metal-poor ones and metal-rich ones.
In the case of the Milky Way, the metal-poor sample ([Fe/H]$<-1.0$) is about three times larger than the metal-rich sample ([Fe/H]$>-1.0$) so the entire sample is dominated by the metal-poor sample. \citet{dic06} show that the turnover magnitude of the metal-poor sample, $M_V = -7.72\pm0.10$, is 
0.34 mag and 0.05 mag brighter than the value for the metal-rich sample and the combined sample in the Milky Way.
Thus the metal-poor sample calibration leads to a distance modulus that is only 0.05 mag larger than the value based on the entire sample calibration. On the other hand, the metal-rich sample calibration  leads to a distance modulus that is 0.29 mag smaller than that based on the entire sample calibration. Similarity of the luminosity of NGC 4993 and the Milky Way
indicates that the globular cluster system in NGC 4993 may be dominated by the metal-poor globular clusters.

In summary, the realistic systematic error for the distance estimate in this study is estimated to be $<0.3$ mag.

\section{Discussion}

\subsection{Comparison with Previous Fundamental Plane Distance Estimates}

We compare our distance measurement with previous 
estimates for NGC 4993 based on the fundamental plane 
in the literature.
\citet{im17} determined the distances from  multi-band data ($BVRIJHK$) for NGC 4993, adopting 
seventeen parameter sets of the fundamental plane in the literature. 
The zeropoint of these calibrations depends on the value of the Hubble constant: $d_A= (R_{\rm eff} h_{70}^{-1} / {\rm kpc}) \times {206265 / {\theta_{\rm eff} /{\rm arcsec}} }$ where $R_{\rm eff}$ and $\theta_{\rm eff}$ are linear and angular effective radii of galaxies, respectively. 
Adopting a value of $H_0=70$ \kmsMpc, 
\citet{im17} obtained a mean value of the angular diameter distance, $d_A= 37.7\pm8.7$ Mpc (corresponding to a luminosity distance of $d_L = 38.4\pm8.9$ Mpc). 

Independently \citet{hjo17} used F606W($V$)-band data for NGC 4993, adopting the calibration of the fundamental plane based on the Leo I Group in \citet{hjo97}. 
They presented 
a luminosity distance, $44.0\pm7.5$ Mpc, which is $\sim$6 Mpc larger than the result given by \citet{im17}. The error in this estimate is dominated by the scatter in the fundamental plane relation (about 17\%) \citep{jor96}.
The zeropoint of the calibration 
in \citet{hjo97} is based on the Cepheid distance to M96 
($d=11.3\pm0.09$ Mpc), one of the spiral galaxies in the Leo I Group. This value is similar to the recent estimate for M96 based on the tip of the red giant branch (TRGB) \citep{lee93} given by \citet{jan17}, $d=11.08\pm0.28$ Mpc. 
However, all four galaxies used for deriving the fundamental plane relation of the Leo I Group are elliptical or lenticular galaxies, and the spiral galaxy M96 was used only for adopting the distance to the Leo I Group. It will be better to use the distances to the same galaxies as used for deriving the fundamental plane relation. 
M105 (NGC 3379), as well as NGC 3377, is 
the brightest member among the Leo I Group galaxies used for the study of the fundamental plane in \citet{hjo97}.
 Recently \citet{lee16b} presented a TRGB distance to M105 derived from deep HST images: 
$d=10.23\pm0.09$ Mpc ($(m-M)_0=30.05\pm0.12$). 
The distance to M105 is about 10\% smaller than the value for M96 adopted in \citet{hjo97}. If we adopt this value as a distance to the Leo I Group to calibrate the zeropoint of the fundamental plane relation, we obtain a 10\% smaller distance for NGC 4993 than the value in \citet{hjo17}. 

Thus the GCLF distance for NGC 4993 in this study, 
$d=41.65\pm3.00$ Mpc, 
is consistent, within the errors, with the values based on the fundamental plane relation.

\subsection{Implication for the Gravitational-wave Standard Siren Method}

Applying the gravitational-wave standard siren method \citep{sch86} to GW170817, \citet{abb17b} derived simultaneously the values of the distance, binary inclination angle, and the Hubble constant: $d=43.8^{+2.9}_{-6.9}$ Mpc, $i=167^{+13}_{-23}$ deg, and
$H_0=70.0^{+12.0}_{-8.0}$ \kmsMpc for
maximum a posteriori (MAP) intervals in their models
(MAP value and smallest range including 90\% of the posterior), 
which were presented as representative values in their paper.
They also derived slightly different values for a symmetric interval (e.g., median and 5\% to 95\% range):
$d=41.1^{+4.0}_{-7.3}$ Mpc, $i=152^{+14}_{-17}$ deg, and
$H_0=74.0^{+16.0}_{-8.0}$ \kmsMpc.

The GCLF distance for NGC 4993 in this study 
is in excellent agreement with the values based on the gravitational wave method (being closer to the value for a symmetric interval  in \citet{abb17b}).
This is remarkable, considering that the two methods are entirely independent in deriving a galaxy distance.
This shows that the gravitational wave method
has a great potential for deriving a precise estimation of the Hubble constant, once the sample of the gravitational wave detection is large enough in the future.
If we use an independent measurement of the distance in the analysis of the gravitational-wave standard siren method, we can determine more tightly the values of the inclination angle and the Hubble constant.



\bigskip
This work was supported by the National Research Foundation of Korea (NRF) grant
funded by the Korean Government (NRF-2017R1A2B4004632).
J.K. was supported by the Global Ph.D. Fellowship Program (NRF-2016H1A2A1907015) of the National Research Foundation.
M.I. was supported by the National Research Foundation of Korea (NRF) grant
(NRF-2017R1A3A3001362).

\clearpage



\clearpage


\end{document}